\documentclass[a4paper,12pt]{article}
\usepackage[pctex32]{graphics}
\usepackage{amsmath,amssymb}
\usepackage{multirow}
\usepackage{latexsym}
\usepackage{epsfig}
\usepackage[english]{babel}
\newcommand{\be}{\begin{equation}}
\newcommand{\ee}{\end{equation}}
\newcommand{\ba}{\begin{eqnarray}}
\newcommand{\ea}{\end{eqnarray}}
\newcommand{\nn}{\nonumber}
\numberwithin{equation}{section} \oddsidemargin 0 mm \evensidemargin
0 mm \topmargin -10 mm \textheight 215 mm \textwidth 163 mm

\renewcommand{\thefootnote}{\fnsymbol{footnote}}
\begin{document}

\begin{titlepage}

\vspace{5mm}
\begin{center}
{\Large \bf Quasinormal frequencies of a massless scalar \\in the
hidden Kerr/CFT proposal}

\vskip .6cm

\centerline{\large Taeyoon Moon$^{a}$ and Yun Soo Myung$^{b}$}

\vskip .6cm
 {Institute of Basic Science and Department of Computer Simulation, \\Inje University, Gimhae 621-749, Korea \\}

\end{center}

\begin{center}

\underline{Abstract}
\end{center}
The hidden Kerr/CFT proposal implies that the massless scalar wave
equation in the near-region and low-frequency limit respects a
hidden SL(2,R)$\times$SL(2,R) invariance in the Kerr black hole
spacetime. We may use this symmetry to determine quasinormal
frequencies (QNFs) of the massless scalar wave propagating around
the Kerr black hole algebraically. It is shown that QNFs obtained
using the hidden conformal symmetry near the horizon correspond
approximately to not only those of scalar perturbation around the
near-horizon region of a nearly extremal Kerr (NEK) black hole, but
also  those of non-equatorial scalar modes around the NEK black
hole. This indicates that the hidden Kerr/CFT proposal could
determine quasinormal modes and frequencies of the massless scalar
wave around the NEK (rapidly rotating) black hole approximately.

\vskip .6cm

 \vskip 0.8cm

\vspace{15pt} \baselineskip=18pt

\noindent $^a$tymoon@sogang.ac.kr\\
\noindent $^b$ysmyung@inje.ac.kr

\thispagestyle{empty}
\end{titlepage}
\renewcommand{\thefootnote}{\arabic{footnote}}
\setcounter{footnote}{0}

\newpage

\section{Introduction}

It was known that  the scalar wave equation in the near-region and
low-frequency limit enjoys a hidden conformal symmetry  in the
generic Kerr black hole which is not regarded as  an underlying
symmetry of the Kerr spacetime
itself~\cite{Castro:2010fd,Krishnan:2010pv}. Such a hidden symmetry
originates from the observation that low-frequency scattering
amplitudes of scalar from  a black hole are expressed  in terms of
hypergeometric functions \cite{Cvetic:1997xv,Cvetic:2011hp}, which
form representations of the conformal group SL(2,R). Curiously, this
might lead to the conjecture that the non-extremal Kerr black hole
with angular momentum $J$ is dual to a conformal field theory (CFT)
with central charges~\cite{Castro:2010fd}.  It is also found that
the low-frequency  amplitudes of scalar-Kerr scattering coincide
with thermal correlators of a CFT on the boundary.

However, we note that this hidden Kerr/CFT proposal should be
contrasted with the usual geometric approach where Kerr/CFT
correspondence may be applied for (nearly) extremal Kerr black hole
with infinite throat geometries containing AdS
subspaces~\cite{Guica:2008mu,Compere:2012jk}. For example, the
near-horizon geometry of the extremal Kerr black hole ($a=M$)
contains  AdS$_2$, whose isometry is SL(2,R)$_R\times$U(1)$_L$. In
this sense, one has to use the terminology of ``hidden" for the
generic Kerr black hole spacetime when taking account of the
near-region $\omega r \ll 1$ and low-frequency limit $\omega M \ll
1$. We would like to point out that the Cardy formula  matches the
Bekenstein-Hawking entropy of the non-extremal Kerr black hole only
when using  the central charges of $c_L=c_R=12J$ obtained from the
Kerr/CFT correspondence for the extremal Kerr black hole.  This is a
weak point of the hidden Kerr/CFT proposal to obtain the generic
Kerr black hole entropy, since it  assumed that the central charges
behave smoothly and they do not change as we move away from the
extremal case~\cite{Bredberg:2011hp}.

On the other hand, Chen and Long~\cite{Chen:2010ik} have proposed
that one can use the hidden conformal symmetry to determine
quasinormal modes (QNMs) algebraically as descendants of a highest
weight state in the black hole spacetime (see also the recent works
\cite{Castro:2013lba,Cvetic:2013lfa} to find QNMs in different
method). Using this idea, the authors~\cite{Bertini:2011ga} have
argued that the scalar wave equation in the near-region and
low-energy limit enjoys a hidden SL(2,R) invariance in the
Schwarzschild spacetime. They have used the SL(2,R) symmetry to
determine algebraically quasinormal frequencies (QNFs:
$\omega=\omega_R-i\omega_I$) of scalar around the Schwarzschild
black hole, leading to  purely imaginary QNFs ($\omega=-i\omega_I$)
which may be suitable for describing  large damping.  Here we note
that the sign of $\omega_I$ is important because it determines the
stability of the black hole~\cite{Berti:2009kk}.  We wish to mention
that the method developed for the hidden Kerr/CFT correspondence
could not be directly applied to derive QNFs of scalar around the
Schwarzschild black hole because there is no room to accommodate an
AdS$_2$ structure in the near-horizon geometry of Schwarzschild
black hole~\cite{Kim:2012ax}. Its near-horizon geometry is the
Rindler spacetime. Furthermore, it seems that no purely imaginary
QNFs were found from the scalar propagating around the Schwarzschild
black hole~\cite{Berti:2009kk,Konoplya:2011qq}.

Recently, the authors~\cite{Ortin:2012mt} have developed  a hidden
SL(2,R) symmetry in the near-region and low-energy limit of the
Reissner-Nordstr\"om (RN) black hole. One may use the SL(2,R)
symmetry to determine QNFs of scalar around the generic RN black
hole algebraically.   QNMs and purely imaginary QNFs were obtained
by employing the operator method~\cite{Kim:2012mh}. As is well
known, QNMs are determined by solving a scalar wave equation around
the RN black hole as well as imposing the boundary conditions:
ingoing waves at the horizon and outgoing waves at infinity of
asymptotically flat spacetime.    A key point in deriving QNMs is to
impose proper boundary conditions on waves. It is not possible  to
derive QNMs of a scalar propagating in the RN black hole spacetime
by using  hidden conformal symmetry solely because  QNMs do not
satisfy the outgoing wave-boundary condition required in
asymptotically flat spacetime. Instead, it is suggested  that QNMs
satisfy the ingoing-boundary condition at the horizon and Dirichlet
boundary condition at infinity.   Hence, we have to find a specified
RN black hole spacetime where one could capture QNMs and QNFs of
scalar in the whole spacetime outside black hole.  Explicitly, in
order to obtain QNMs and QNFs of scalar around the RN black hole
imposed by the near-region and low-frequency limit, we have to
consider a freely  propagating scalar around the corresponding
(specified) black hole spacetime  without requiring the near-region
and low-frequency limit.
 It is known that the
specified   RN black hole spacetime is given by the near-horizon
region AdS$_2\times S^2$ of a nearly extremal RN black
hole~\cite{Chen:2012zn}.  This means that in order to derive QNMs
and QNFs of scalar around black hole with hidden conformal symmetry,
the SL(2,R) hidden symmetry should be realized as an isometry
symmetry of the specified RN black hole.
 Also, it is worth to note
that purely imaginary QNFs of a  scalar reflects lower-dimensional
nature of AdS$_2$ base in the near-horizon region of a nearly
extremal RN black hole~\cite{Kim:2012mh,Cordero:2012je}. This
implies that purely imaginary QNFs  could not be derived from scalar
 around any RN black holes in four-dimensional
spacetime~\cite{Berti:2009kk,Konoplya:2011qq}.

In the case of Kerr black hole spacetime imposed by hidden conformal
symmetry~\cite{Lowe:2011wu,Lowe:2011aa}, the specified black hole
spacetime is conjectured to be the near-horizon region of a nearly
extremal Kerr (NEK, rapidly rotating) black hole with $a\to M$
because the latter contains   AdS$_2$ base as the near-horizon
region.
 In this sense, we
expect to obtain   QNFs of scalar around the Kerr black hole with
hidden conformal symmetry.

In four-dimensional spacetime, NEK black holes have considerable
theoretical and observational significance. For a rapidly rotating
stellar-mass black hole, its spin measurement was reported
in~\cite{McClintock:2006xd,McClintock:2009as}, while for a rapidly
rotating supermassive black  hole, its spin measurement was very
recently reported in~\cite{Risaliti:2013cga}. Detweiler has made an
approximation to the Teukolsky
equation~\cite{Teukolsky:1972my,Teukolsky:1974yv} for NEK black
holes to show that QNMs with $\ell=m$ (equatorial modes) have a long
decay time of $1/\omega_I$~\cite{Detweiler:1980gk}.  Sasaki and
Nakamura have computed QNFs analytically~\cite{Sasaki:1989ca}, and
Anderson and Glampedakis have proposed long-lived emissions from NEK
black holes~\cite{Andersson:1999wj}.  However, it seems that there
exists a long-standing controversy about what set of QNMs decay
slowly~\cite{Hod:2008zz} and whether $\omega_I$ vanishes as $a \to
M$~\cite{Cardoso:2004hh}. Here we will use analytic expressions of
complex QNFs~\cite{Hod:2008zz,Yang:2012pj} of scalar around  NEK
black holes to compare  QNFs of Kerr black hole with the hidden
conformal symmetry.

We will show that  QNFs obtained using the hidden conformal symmetry
near the horizon correspond approximately to not only those of
scalar perturbation around the near-horizon region of a nearly
extremal Kerr (NEK) black hole, but also  those of non-equatorial
($\ell\not= m$) scalar modes around the NEK black hole. However,
this approach could not determine QNFs of equatorial $(\ell=m$)
modes around the NEK black hole.  This indicates that the hidden
Kerr/CFT proposal could determine quasinormal modes and frequencies
of scalar around the NEK (rapidly rotating) partly.

\section{Hidden conformal symmetry in Kerr geometry}

Let us consider the Kerr metric in Boyer-Lindquist coordinates,
whose form is given by
\begin{eqnarray}\label{ker}
ds^2=-\frac{\triangle}{z^2}\Big(dt-a\sin^2\theta
d\phi\Big)^2+\frac{z^2}{\triangle}dr^2+\frac{\sin^2\theta}{z^2}
\Big((r^2+a^2)d\phi-adt\Big)^2+z^2d\theta^2,
\end{eqnarray}
where $\triangle$, $z^2$, and $a$ denote
\begin{eqnarray}
\triangle=r^2-2Mr+a^2,~~~z^2=r^2+a^2\cos^2\theta,~~~a=\frac{J}{M}
\end{eqnarray}
with the black hole mass $M$ and the angular momentum $J$. Two
solutions to  $\triangle=0$ are defined by the inner ($r_-$) and
outer ($r_+$) horizons as
\begin{eqnarray}
r_{\pm}=M\pm\sqrt{M^2-a^2}\equiv M\pm
r_0,~~r_0=\sqrt{M^2-a^2}=\frac{r_+-r_-}{2}.
\end{eqnarray}
It is well-known that in the Kerr geometry (\ref{ker}), the massless
Klein-Gordon equation $\nabla_{\mu}\nabla^{\mu}\Phi=0$ for the
ansatz $\Phi(t,r,\theta,\phi)=e^{-i\omega t+im\phi}R(r)S(\theta)$
can be separated as
\begin{eqnarray}\label{theta}
\left[\frac{1}{\sin\theta}\partial_{\theta}(\sin\theta\partial_{\theta})
-\frac{m^2}{\sin^2\theta}+\omega^2a^2\cos^2\theta-K_{\ell}\right]S(\theta)=0
\end{eqnarray}
and
\begin{eqnarray}\label{oradial}
&&\left[\partial_{r}(\triangle\partial_r)+\frac{(2Mr_+\omega-am)^2}{2r_0(r-r_+)}
-\frac{(2Mr_-\omega-am)^2}{2r_0(r-r_-)}-K_{\ell}\right.\nn\\
&&\left.\hspace*{16.5em}+\Big(r^2+2M(r+2M)\Big)\omega^2\right]R(r)=0
\end{eqnarray}
with $K_{\ell}$ the separation constant to be fixed below. We note
that Eq.(\ref{oradial}) is exactly the same with the radial
Teukolsky equation \cite{Teukolsky:1972my,Teukolsky:1974yv} for
$s=0$. In order to investigate the hidden conformal
symmetry~\cite{Castro:2010fd}, we first consider the low-frequency
limit  $\omega M\ll1$ and near-region $\omega r\ll1$. It turns out
that in the low-frequency limit of $\omega M\ll1$, Eq.(\ref{theta})
reduces to
\begin{eqnarray}\label{theta0}
\left[\frac{1}{\sin\theta}\partial_{\theta}(\sin\theta\partial_{\theta})
-\frac{m^2}{\sin^2\theta}\right]S(\theta)=K_{\ell}S(\theta),
\end{eqnarray}
which corresponds to the spherical Laplacian that can be solved in
terms of spherical harmonic function
$Y_{\ell}^{m}(\theta,\phi)=e^{im\phi}S(\theta)$ for
$K_{\ell}=\ell(\ell+1)$. On the other hand, the radial equation
(\ref{oradial}) in the limits of  $\omega r\ll1$ and $\omega M\ll1$
becomes
\begin{eqnarray}
\label{radial0}
\left[\partial_{r}(\triangle\partial_r)+\frac{(2Mr_+\omega-am)^2}{2r_0(r-r_+)}
-\frac{(2Mr_-\omega-am)^2}{2r_0(r-r_-)}\right]R(r)=\ell(\ell+1)R(r),
\end{eqnarray}
where the last term in Eq. (\ref{oradial}) is dropped and we set
$K_{\ell}=\ell(\ell+1)$. It is known that (\ref{radial0}) could  be
converted into hypergeometric equation whose solution is given by
hypergeometric functions.  These form representations of SL(2,R),
indicating the existence of  hidden conformal symmetry. To show the
hidden conformal symmetry explicitly, we introduce two sets of
vector fields \cite{Castro:2010fd}
\begin{eqnarray}
\label{oh1}H_1&=&ie^{-2\pi
T_R\phi}\Big(\triangle^{1/2}\partial_r+\frac{1}{2\pi
T_R}\frac{r-M}{\triangle^{1/2}}\partial_{\phi}
+\frac{2T_L}{T_R}\frac{Mr-a^2}{\triangle^{1/2}}\partial_t\Big),\\
H_0&=&\frac{i}{2\pi T_R}\partial_{\phi}+2iM\frac{T_L}{T_R}\partial_t,
\label{h0}\\
H_{-1}&=&ie^{2\pi
T_R\phi}\Big(-\triangle^{1/2}\partial_r+\frac{1}{2\pi
T_R}\frac{r-M}{\triangle^{1/2}}\partial_{\phi}
+\frac{2T_L}{T_R}\frac{Mr-a^2}{\triangle^{1/2}}\partial_t\Big)
\end{eqnarray}
and
\begin{eqnarray}
\bar{H}_1&=&ie^{-2\pi T_L\phi +(t/2M)}\Big(\triangle^{1/2}\partial_r
-\frac{a}{\triangle^{1/2}}\partial_{\phi}-2M\frac{r}{\triangle^{1/2}}\partial_t\Big),\\
\bar{H}_0&=&-2iM\partial_t,\label{bh0}\\
\label{obhm}\bar{H}_{-1}&=&ie^{2\pi T_L\phi
-(t/2M)}\Big(-\triangle^{1/2}\partial_r
-\frac{a}{\triangle^{1/2}}\partial_{\phi}-2M\frac{r}{\triangle^{1/2}}\partial_t\Big),
\end{eqnarray}
which satisfy the SL(2,R) algebra
\begin{eqnarray}\label{alg}
[H_0,H_{\pm1}]=\mp i H_{\pm1},~~~~~[H_{-1},H_{1}]=-2i H_0,
\end{eqnarray}
and similarly for $\bar{H}_0$, $\bar{H}_{\pm1}$. Here the left/right
temperatures of CFT are given by
\begin{eqnarray}
T_L=\frac{r_++r_-}{4\pi a},~~~T_R=\frac{r_0}{2\pi a},
\end{eqnarray}
while the Hawking temperature of the generic Kerr black hole takes
the form \be T_{\rm H}=\frac{r_0}{4\pi M r_+}=T_R\Omega \ee  with
the angular velocity of the black hole at the horizon \be
\Omega=\frac{a}{2Mr_+}=\frac{a}{r_+^2+a^2}. \ee
 The two Casimir operators ${\cal
H}^2={\cal \bar{H}}^2$, being obtained from the above vector fields
can be written as
\begin{eqnarray}
{\cal H}^2&=&-H_0^2+\frac{1}{2}(H_1H_{-1}+H_{-1}H_1),\label{H2}\\
{\cal \bar{H}}^2&=&-\bar{H}_0^2+\frac{1}{2}(\bar{H}_1\bar{H}_{-1}
+\bar{H}_{-1}\bar{H}_1)\label{bH2}\\
&=&\partial_{r}(\triangle\partial_r)-\frac{(2Mr_+\partial_t-a\partial_\phi)^2}{2r_0(r-r_+)}
+\frac{(2Mr_-\partial_t-a\partial_\phi)^2}{2r_0(r-r_-)},\label{h2bh2}
\end{eqnarray}
which allow the radial equation (\ref{radial0}) to rewrite two
eigenvalue equations as
\begin{eqnarray}\label{hbarh}
{\cal H}^2\Phi={\cal \bar{H}}^2\Phi=\ell(\ell+1)\Phi
\end{eqnarray}
for $\Phi$.

 Consequently, the above result shows that in the
near-region and low-frequency limit, the scalar  equation
(\ref{hbarh}) admits SL(2,R)$_L\times$SL(2,R)$_R$ symmetry with
conformal weights\footnote{These conformal weights are obtained when
one plugs  (\ref{pri}) into  (\ref{H2}) and (\ref{bH2}). Then, one
rearranges  it by making use of  (\ref{alg}) and (\ref{H1}).}
\begin{eqnarray}\label{wei}
(h_L, h_R)=(\ell+1,\ell+1).
\end{eqnarray}

\section{Potential behaviors}

In order to see how much is changed when taking two limits, we wish
to study the change of potential around the Kerr black hole which
feels by a scalar. For this purpose, we define a tortoise coordinate
$r^*$ defined  by $dr^*=dr (r^2+a^2)/\triangle$, \be
r^*=r+\frac{1}{2\kappa}\ln\Big[\frac{r-r_+}{2M}\Big]-\frac{Mr_-}{r_0}\ln\Big[\frac{r-r_-}{2M}\Big],~~\kappa=2\pi
T_{\rm H} \ee where $r\in[r_+,\infty)$ is mapped into
$r^*\in(-\infty,\infty)$. Introducing a radial function
$\psi(r)=R(r)\sqrt{r^2+a^2}$, the original radial equation
(\ref{oradial}) reduces to the Schr\"{o}dinger-type equation
\begin{eqnarray} \label{tteq}
\frac{d^2\psi}{dr^{*2}}+\Big(\omega^2-V_{\rm K}\Big)\psi=0,
\end{eqnarray}
whose potential $V_{\rm K}$ is given by
\begin{eqnarray}\label{K1pots}
V_{\rm K}=\omega^2+\frac{(r-r_+)(r-r_-)}{(r^2+a^2)^2}\Bigg[
\frac{2Mr^3+a^2r^2-4Ma^2+a^4}{(r^2+a^2)^2}-\tilde{K}_{\ell}(r)\Bigg]
\end{eqnarray}
with $\tilde{K}_{\ell}(r)$
\begin{eqnarray}
\tilde{K}_{\ell}(r)=\frac{(2Mr_+\omega-am)^2}{2r_0(r-r_+)}
-\frac{(2Mr_-\omega-am)^2}{2r_0(r-r_-)}
+\Big(r^2+2M(r+2M)\Big)\omega^2-K_{\ell}.
\end{eqnarray}
One can check easily that the scalar potential around the
Schwarzschild black hole is recovered from taking $a=0$ ($r_+=2M$)
in the Kerr potential $V_{\rm K}$ (\ref{K1pots})
\begin{eqnarray} \label{sch}
\hspace*{4.5em}V_{\rm
Sch}^{a=0}=\left(1-\frac{2M}{r}\right)\Bigg[\frac{2M}{r^3}
+\frac{\ell(\ell+1)}{r^2}\Bigg]~~~~~{\rm with}~~
K_{\ell}=\ell(\ell+1).
\end{eqnarray}
On the other hand, for the limits of $\omega r\ll1$ and $\omega
M\ll1$,  the Kerr potential $V_{\rm K}$ (\ref{K1pots}) is
approximated to be
\begin{eqnarray}\label{potss}
V_{\rm K}^{\omega r,\omega M\ll1}=V_{\rm
K}+\frac{(r-r_+)(r-r_-)}{(r^2+a^2)^2}
\Big[r^2+2M(r+2M)\Big]\omega^2.
\end{eqnarray}
\begin{figure*}[t!]
   \centering
   \includegraphics{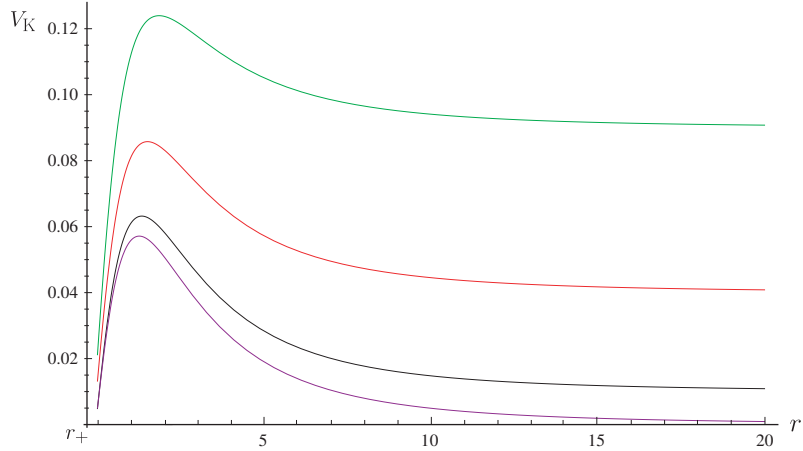}
\caption{Various potentials as functions of $r$ for $M=2/3$,
$a=\sqrt{3}/3$, $m=0.1$, and $r_+=1$. The bottom curve depicts the
Kerr potential $V_{\rm K}(r)$, while three others are $V_{\rm
K}^{\omega r,\omega M\ll1}$ for $\omega=0.1,~0.2,~$and $0.3$ from
bottom to top. For large $r$, it  shows that $V_{\rm K}\to0$, while
$V_{\rm K}^{\omega r,\omega M\ll1}\to\omega^2$.}
\end{figure*}
It is shown that as $r\to\infty$, the potential $V_{\rm K}$
 approaches $0$, while $V_{\rm K}^{\omega r,\omega
M\ll1}$  approaches  $\omega^2$, which implies that its asymptotic
behavior depends on the value of the frequency $\omega$. From Fig.
1, we observe a difference between $V_{\rm K}|_{r\to\infty}\to 0$
and $V_{\rm K}^{\omega r,\omega M\ll1}|_{r\to\infty}\to\omega^2$ for
large $r$. In the near-horizon of $r\approx r_+$, however, two
potentials have the same form as \be V_{\rm K}^{\omega r,\omega
M\ll1}\Big|_{r\approx r_+}=V_{\rm K}\Big|_{r\approx
r_+}=\omega^2-(\omega^2-m\Omega)^2.\ee Their waves are given by
solving (\ref{tteq}) as \be
\psi_+=Ae^{-i(\omega-m\Omega)r^*}+Be^{i(\omega-m\Omega)r^*}.\ee It
should be pointed out that the near-horizon region of $r\approx r_+$
is not enough to derive quasinormal modes because we have to impose
boundary conditions at the horizon
 as well as asymptotic infinity. In
order to describe the near-horizon region well, it is useful to
introduce a new coordinate defined by
\begin{eqnarray}
\rho=-\frac{1}{2\kappa}\ln\left[\frac{r-r_+}{r-r_-}\right],
\end{eqnarray}
where $r\in[r_+,\infty)$ is inversely mapped into
$\rho\in(\infty,0]$. In the near-horizon limit, we have a relation
of $r\approx r_++2r_0 e^{-2\kappa \rho}$ which leads to $r\approx
r_+$ for $\rho(=-r^*) \to \infty$.
 Using
this new coordinate, the radial equation (\ref{oradial}) leads to
the Schr\"odinger equation
\begin{eqnarray} \label{rhoeq}
\frac{d^2}{d\rho^{2}}R(\rho)+\Big(\omega^2-V_{\rm
K}(\rho)\Big)R(\rho)=0,
\end{eqnarray}
where the potential $V_{\rm K}(\rho)$ is given by
\begin{eqnarray}\label{vrho}
V_{\rm K}(\rho)=\omega^2-\frac{\kappa^2}{\sinh^2[\kappa\rho]}
\tilde{K}_{\ell}(\rho)
\end{eqnarray}
with $\tilde{K}_{\ell}(\rho)$
\begin{eqnarray}
&&\hspace*{-5.5em}\tilde{K}_{\ell}(\rho)~=~\frac{\sinh^2[\kappa\rho]}{r_0^2}
\Bigg[\Big((M+r_0\coth[\kappa\rho])^2+a^2\Big)\omega
-am\Bigg]^2\nn\\
&&\hspace*{17em}+~2ma\omega-a^2\omega^2-K_{\ell}.
\end{eqnarray}
We note that in the Schwarzchild limit ($a\to0$), the potential
(\ref{vrho}) reduces to
\begin{eqnarray}
V_{\rm K}^{a\to 0}(\rho)\to\omega^2
\Bigg[1-\frac{1}{16e^{-4\kappa\rho}\sinh^4[\kappa\rho]} \Bigg]
+\frac{\ell(\ell+1)\kappa^2}{\sinh^2[\kappa\rho]},
\end{eqnarray}
which is exactly the same as found in Ref.\cite{Kim:2012ax}. On the
other hand, our main concern is to take the low-frequency and
near-horizon\footnote{Note that ``near-horizon'' region is not the
same as the near-region for given geometry. This is because $r$ in
the near-region could  be arbitrarily large for a sufficiently small
$\omega$.} limits in (\ref{vrho}), which yields a simple potential
form (see Appendix A for explicit derivation)
\begin{eqnarray}\label{vkom}
V_{\rm K}^{\omega
M\ll1}(\rho)~\approx~\Big[\omega^2-(\omega-m\Omega)^2\Big]
+\frac{\ell(\ell+1)\kappa^2}{\sinh^2[\kappa\rho]}.
\end{eqnarray}
Here, the first term of (\ref{vkom}) appears as a new term when
comparing $V_{\rm K}^{\omega M\ll1}$ with those for the
Schwarzschild black hole~\cite{Kim:2012ax} and the RN black
hole~\cite{Kim:2012mh}. Note that this term disappears for either
the non-rotating black hole ($a=0$) or axisymmetric perturbation
($m=0$)~\cite{Compere:2012jk}.

Consequently, in the low-frequency and near-horizon limits the
Eq.(\ref{rhoeq}) together with (\ref{vkom}) leads to the
Schr\"odinger-like equation
\begin{eqnarray} \label{rrhoeq}
\frac{d^2}{d\rho^{2}}R(\rho)+\Big[(\omega-m\Omega)^2-V_{\rm
HCS}(\rho)\Big]R(\rho)=0,
\end{eqnarray}
where the HCS\footnote{We would like to clarify a terminology of
HCS. Here, the hidden conformal symmetry (HCS) has  nothing to do
with ${\cal H}$ and ${\cal\bar{H}}$ (\ref{H2})-(\ref{h2bh2}) given
in the near-region limit, but it is related to ${\cal H}_{\rho}$
(\ref{rH2}) obtained by taking the near-horizon as well as
near-region limits.}-potential $V_{\rm HCS}$ is given
by~\cite{Kim:2012ax,Kim:2012mh} \be V_{\rm
HCS}=\frac{\ell(\ell+1)\kappa^2}{\sinh^2[\kappa\rho]}.\ee It should
be pointed out that Eq. (\ref{rrhoeq}) is the same forms obtained
for the Schwarzschild and RN black holes when replacing
$(\omega-m\Omega)^2$ by $\omega^2$, which implies that the potential
$V_{\rm HCS}$
 behaves as the potential of a scalar field around the AdS black
hole because  $V_{\rm HCS}\to\infty$ at asymptotic infinity
($\rho\to0$). In the next section, we will use Eq. (\ref{rrhoeq}) to
construct the new SL(2,R) three vector fields based on the
$\rho$-representation for the Kerr black hole.

\section{Quasinormal frequencies by operator method}

First, we wish to derive  QNFs of scalar around the Kerr black hole
by employing an  algebraic method based on  the hidden conformal
symmetry (\ref{oh1})-(\ref{obhm})~\cite{Lowe:2011wu}. For
convenience, we identify $L_n=-iH_n$ and $\bar{L}_n=-i\bar{H}_n$, so
that the $L_n$'s and $\bar{L}_n$ satisfy the Witt algebra,
respectively, \be
[L_n,L_m]=(n-m)L_{n+m},~~[\bar{L}_n,\bar{L}_m]=(n-m)\bar{L}_{n+m}.
\ee We first consider the primary state $\Phi^{(0)}$ with the
conformal weight $h_L$ and $h_R$
\begin{eqnarray}\label{pri}
L_0\Phi^{(0)}=h_L\Phi^{(0)},~~\bar{L}_0\Phi^{(0)}=h_R\Phi^{(0)}.
\end{eqnarray}
Acting operations $L_0$ and $\bar{L}_0$  on $\Phi^{(0)}$ with the
ansatz
\begin{eqnarray}
\Phi^{(0)}=e^{-i\omega_0 t+im\phi}R^{(0)}(r)S(\theta),
\end{eqnarray}
then equation (\ref{pri}) yields the relation  between $\omega$ and
$m$ as
\begin{eqnarray}\label{om}
\omega_0=\frac{am}{2Mr_+}=m \Omega,
\end{eqnarray}
when requiring  the same conformal weight condition $h_L=h_R$. In
this case, $h_L$ and $h_R$ are determined to be
\begin{eqnarray}\label{hlr}
h_L=h_R=\frac{iam}{r_+}=2iM \omega_0.
\end{eqnarray}
As a result,  from (\ref{wei}) and (\ref{hlr}), the frequency
$\omega_0$ is given by purely imaginary quantity
\begin{eqnarray}
\omega_0=-i\frac{\ell+1}{2M}
\end{eqnarray}
and also, the azimuthal number $m$ becomes purely imaginary quantity
 \be m=\frac{\omega_0}{\Omega}=-i 2\pi (T_R+T_L)(\ell+1).
\ee  In addition, the highest weight condition is implemented  by
requiring two conditions on scalar $\Phi^{(0)}$
\begin{eqnarray}\label{H1}
L_1\Phi^{(0)}=\bar{L}_1\Phi^{(0)}=0.
\end{eqnarray}
It turns out that  the solution to (\ref{H1}) is given by
\begin{eqnarray}\label{Phi0}
\Phi^{(0)}(t,r,\phi)~=~C\Big(r^2+2(M-r)r_--a^2\Big)^{-i\frac{m}{4\pi(T_R+T_L)}}e^{-i\omega_0
t+im\phi}
\end{eqnarray}
where $C$ is an integration constant. Staring from the highest
weight  state $\Phi^{(0)}$, we may construct all descendants
$\Phi^{(n)}$ by using the relations
\begin{eqnarray}
\Phi^{(n)}&=&e^{-i\omega_n
t+im_n\phi}R^{(n)}(r)S(\theta)\nn\\
&=&(\bar{L}_{-1})^n\Phi^{(0)}.
\end{eqnarray}
$\omega_n,~m_n$, and $R^{(n)}$ are determined to give
\begin{eqnarray}
\omega_n&=&\omega_0-i\frac{n}{2M}~=~-i\frac{1}{2M}(\ell+1+n),\label{qnfs} \\
m_n&=&-i 2\pi T_R(\ell+1)-i2\pi T_{\rm L}(\ell+1+n), \label{qnms}\\
R^{(n)}(r)&=&\triangle^{-\frac{n}{2}}\Big(-\triangle\partial_r-iam_{n-1}
+i2Mr\omega_{n-1}\Big)\nn\\
&&\times\triangle^{-\frac{n}{2}}\Big(-\triangle\partial_r-iam_{n-2}
+i2Mr\omega_{n-2}\Big)\nn\\
&&\times\cdots\times\triangle^{-\frac{n}{2}}\Big(-\triangle\partial_r-iam_{0}
+i2Mr\omega_{0}\Big)R^{(0)}(r),
\end{eqnarray}
where $R^{(0)}(r)$ is given by
\begin{eqnarray}
R^{(0)}(r)~=~C\Big(r^2+2(M-r)r_--a^2\Big)^{-i\frac{m}{4\pi(T_R+T_L)}},
\end{eqnarray}
which can be read off from (\ref{Phi0}).  Even though we obtain QNFs
and QNMs from $L_n$ and $\bar{L}_n$, their results are not promising
because neither Eq. (\ref{qnfs}) is usual complex QNFs nor
Im$[\omega_n]\propto 2\pi T_H$, compared with the QNFs
\cite{Hod:2008zz} obtained for near extremal Kerr black holes (see
Appendix B).

Now we turn to derive QNFs and QNMs of scalar around the Kerr black
hole by using (\ref{rrhoeq}), which results from taking near-horizon
as well as low-frequency limit. This suggests that QNFs and QNMs of
scalar around the Kerr black hole may be obtained when replacing
$\omega^2$ by $(\omega-m\Omega)^2$ in the Schwarzschild and RN black
holes. It is straightforward to derive QNFs by introducing three
vector fields
 \ba\label{vecops}
 L_1 &=& \frac{1}{\kappa} e^{\kappa t}
       \Big[\cosh\left(\kappa\rho\right)\partial_t+\sinh\left(\kappa\rho\right) \partial_\rho\Big]\ , \nonumber \\
 L_0 &=& -\frac{1}{\kappa}\partial_t \ , \\
 L_{-1} &=& \frac{1}{\kappa}e^{-\kappa t}
       \Big[\cosh\left(\kappa\rho\right)\partial_t-\sinh\left(\kappa\rho\right) \partial_\rho\Big]\  . \nonumber
 \ea
 These satisfy  the SL(2,R) commutation relations
 \be
  \left[L_0, L_{\pm 1}\right]=\mp L_{\pm 1},
  ~~~\left[L_1, L_{-1}\right]=2L_0.
 \ee
Then, the SL(2,R) Casimir operator is constructed  by
 \ba\label{rH2}
 {\cal H}_\rho^2&=&L^2_0-\frac{1}{2}(L_1L_{-1}+L_{-1}L_1)\nonumber\\
           &=&-\Big(\frac{\sinh(\kappa\rho)}{\kappa}\Big)^2\partial^2_t
              +\Big(\frac{\sinh(\kappa\rho)}{\kappa}\Big)^2\partial^2_\rho   \ .
 \ea
Eq.(\ref{rrhoeq}) could be written as \be \label{effeq} {\cal
H}_\rho^2\tilde{\Phi}(\rho)=\ell(\ell+1)\tilde{\Phi}(\rho), \ee
where $\tilde{\Phi}$ can be effectively written as\be \label{effphi}
\tilde{\Phi}(t,\rho,\theta)\sim
e^{-i(\omega_0-m\Omega)t}R(\rho)S(\theta). \ee First of all, we
define the primary state by $\Phi^{(0)}$ which satisfies
 \be
 L_0\tilde{\Phi}^{(0)}=h\tilde{\Phi}^{(0)},
 \ee
and the highest weight condition
 \be\label{hwc}
 L_1\tilde{\Phi}^{(0)}=0.
 \ee
Considering (\ref{effphi}),  one has a conformal weight
 \be\label{h1}
 h=i\frac{\omega_0-m\Omega }{\kappa}=i \frac{\omega_0-m\Omega}{2\pi T_H}.
 \ee
On the other hand, for  $\tilde{\Phi}^{(0)}$, the SL(2,R) Casimir
operator satisfies
 \be \label{caop}
 {\cal H}^2\tilde{\Phi}^{(0)}=h(h-1)\tilde{\Phi}^{(0)}.
 \ee
Comparing Eq. (\ref{caop}) with Eq. (\ref{effeq}), one has
 \be
 h=\frac{1}{2}[1\pm(2\ell+1)].
 \ee
Together with Eq.~(\ref{h1}), for $h>0$,  one can find
 \be
 \omega_0=m\Omega -i2\pi T_H (\ell+1).
 \ee
Then,  all the descendants are constructed  by
 \be
 \tilde{\Phi}^{(n)}=(L_{-1})^n\tilde{\Phi}^{(0)}
 \ee
so that we have
 \be
\tilde{ \Phi}^{(n)}\equiv e^{-i(\omega_n-m\Omega) t}R^{(n)}(\rho).
 \ee
Here we read off  QNFs  as
 \be \label{impq}
 \omega_n=\omega_0-i\kappa n=m\Omega-i2\pi T_H\Big[n+\ell+1\Big],
 \ee
which is one of our main results.

Moreover, the $n$-th radial eigenfunction $ R^{(n)}(\rho)$ takes the
form
 \ba
 R^{(n)}(\rho)&=&\left(\kappa\right)^{-n}
    \left(-i(\omega_{n-1}-m\Omega)\cosh\left(\kappa\rho\right)-\sinh\left(\kappa\rho\right)\frac{d}{d\rho}\right)\nonumber\\
    &&~\times\left(-i(\omega_{n-2}-m\Omega)\cosh\left(\kappa\rho\right)-\sinh\left(\kappa\rho\right)\frac{d}{d\rho}\right)\nonumber\\
    &&~\cdot\cdot\cdot \times \left(-i(\omega_0-m\Omega)\cosh\left(\kappa\rho\right)-\sinh\left(\kappa
    \rho\right)\frac{d}{d\rho}\right)R^{(0)}(\rho). \label{qnmss}
 \ea
We also have
 \be
 L_0\tilde{\Phi}^{(n)}=(h+n)\tilde{\Phi}^{(n)},
 \ee
which implies that $\tilde{\Phi}^{(n)}$  forms a principal discrete
highest weight representation of the SL(2,R). Now  we wish to  solve
the highest weight condition (\ref{hwc}) to determine the highest
weight state $R^{(0)}(\rho)$
 \be
 \Big[-i(\omega_0-m\Omega)\cosh\left(\kappa\rho\right)+\sinh\left(\kappa\rho\right)\frac{d}{d\rho}\Big]R^{(0)}(\rho)=0.
 \ee
The solution is given by
 \be\label{R0sol}
 R^{(0)}(\rho)=C\Big[\sinh\left(\kappa\rho\right)\Big]^{i\frac{\omega_0-m\Omega}{\kappa}}.
 \ee
We note that the solution (\ref{R0sol}) behaves as
 \be
 R^{(0)} \sim e^{-i(\omega_0-m\Omega) r_*}~~~~~{\rm for}~~r\rightarrow r_+.
 \ee
This is  the ingoing mode propagating into the horizon ($\rho \to
\infty$).  For the $n$-th radial eigenfunction, one can easily show
by induction
 \be
 R^{(n)} \sim e^{-i (\omega_n-m\Omega) r_*}, ~~{\rm as}~ r_* \to -\infty.
 \ee

Finally, we observe that $R^{(0)}(0)=0$ at infinity
($\rho\rightarrow 0, r_* \to \infty)$, which shows that it is not
the outgoing wave at infinity  but satisfies the Dirichlet boundary
condition as like at the infinity of AdS spacetime.  Moreover, the
first radial eigenfunction $R^{(1)}(\rho)$ can be explicitly
constructed as
 \be
 R^{(1)}(\rho) =
  -2iC(\omega_0-m\Omega)\cosh(\kappa\rho)
  \Big[\sinh\left(\kappa\rho\right)\Big]^{i\frac{\omega_0-m\Omega}{\kappa}},
 \ee
which also satisfies the Dirichlet boundary condition at infinity.
One can easily show that the $n$-th radial eigenfunction $
R^{(n)}(\rho)$ behaves as the same way as $R^{(1)}(\rho)$ likewise.

\section{Near-horizon geometry of the NEK black hole}

In the case of Kerr black hole spacetime imposed by hidden conformal
symmetry, it is conjectured that the specified black hole spacetime
could be the near-horizon region of a NEK black hole with $a\to M$.
In this section we wish to study the near-horizon geometry of the
NEK black hole and obtain QNFs of scalar propagating around this
background geometry. For this purpose,  we first consider the NEK
black hole by considering the conditions of $r_0\ll1,~a\approx M,$
and $r_+\approx r_-\approx M$ which are explicitly expressed by
introducing a very small parameter $\lambda$
as~\cite{Bredberg:2009pv}
\begin{eqnarray}
r_\pm &=&M \pm r_0 \approx M \pm  2\pi M \tilde{T}_R \lambda+{\cal O}(\lambda^2)\\
a&=&\sqrt{r_+r_-}\approx M-2M(\lambda\pi \tilde{T}_R)^2+{\cal
O}(\lambda^3),
\end{eqnarray}
which  imply that three temperatures and angular velocity are
approximated to be \be T_L\approx\frac{1}{2\pi},~~T_R\approx
\frac{r_0}{2\pi M}\approx\tilde{ T}_R \lambda,~~T_{\rm NEK}\approx
\frac{r_0}{4\pi M^2}\approx
\frac{\tilde{T}_R\lambda}{2M},~~\Omega\approx \frac{1}{2M}. \ee Here
one is keeping $\tilde{T}_R$  fixed in the limits of $T_{\rm
NEK},T_R \to 0(\lambda\to 0)$ and it is called the dimensionless
near-horizon temperature.
 In order to describe the near-horizon geometry, we also
introduce  three coordinates
\begin{eqnarray} \label{ctran}
\tau=\lambda\frac{t}{2M},~~~y=\frac{r-r_+}{\lambda
r_+},~~~\varphi=\phi-\frac{t}{2M},
\end{eqnarray}
and take $\lambda\to 0$ keeping $(\tau,y,\varphi)$ fixed. After some
manipulations, we obtain  the near-horizon region  of NEK black hole
isolated by the NHEK limit~\cite{Bardeen:1999px}
\begin{eqnarray}
&&\hspace*{-1.5em}ds^2=2J\Gamma(\theta)\Bigg[-y(y+4\pi
\tilde{T}_R)d\tau^2+\frac{dy^2}{y(y+4\pi
\tilde{T}_R)}+d\theta^2+\Lambda^2\Big(d\varphi+(y+2\pi
\tilde{T}_R)d\tau\Big)^2\Bigg],\nn\\
&&\label{nhek}
\end{eqnarray}
where
\begin{eqnarray}
\Gamma(\theta)=\frac{1+\cos^2\theta}{2},
~~~~~\Lambda(\theta)=\frac{2\sin\theta}{1+\cos^2\theta}\nn
\end{eqnarray}
with $\theta\in[0,\pi]$ and $\varphi\sim\varphi+2\pi$. The horizon
is located at $y=0$. For the extremal Kerr black hole, we have
$\tilde{T}_R=0$ and its near-horizon geometry is described by
\begin{eqnarray}\label{ekbh}
ds^2_{\rm
NHEK}=2J\Gamma(\theta)\Bigg[-y^2d\tau^2+\frac{dy^2}{y^2}+d\theta^2+\Lambda^2\Big(d\varphi+y
d\tau\Big)^2\Bigg]
\end{eqnarray}
in terms of Poincare coordinates. This represents the quotient of
warped  AdS$_3\approx$ AdS$_2\times S^1$ for fixed polar angle
$\theta$~\cite{Kim:2009xx}. Hence, the presence of $\tilde{T}_R$
distinguishes between NEK and extremal Kerr black holes.

It is found that in the background spacetime (\ref{nhek}), the
massless Klein-Gordon equation $\nabla_{\mu}\nabla^{\mu}\Phi=0$ with
\be \label{newan}
\Phi(\tau,y,\theta,\varphi)=e^{-i\tilde{\omega}\tau+im\varphi}Y(y)S(\theta)\ee
 can be decomposed into two  differential equations
\begin{eqnarray}\label{theta1}
\left[\frac{1}{\sin\theta}\partial_{\theta}(\sin\theta\partial_{\theta})
-\frac{m^2}{\sin^2\theta}-\frac{m^2}{4}\sin^2\theta+\bar{K}_{\ell
m}\right]S(\theta)=0
\end{eqnarray}
and
\begin{eqnarray}\label{rad1}
\Bigg[\partial_{y}\Big(y(y+4\pi
\tilde{T}_R)\partial_y\Big)-\bar{K}_{\ell
m}+m^2+\frac{\Big(\tilde{\omega}+m(y+2\pi
\tilde{T}_R)\Big)^2}{y(y+4\pi \tilde{T}_R)}\Bigg]Y(y)=0,
\end{eqnarray}
where $\bar{K}_{\ell m}$ are eigenvalues which can  be computed
numerically~\cite{Bardeen:1999px}. Approximately, it is given by
$\bar{K}_{\ell m}\approx \ell(\ell+1)+cm^2$ with $c\in[0.13,0.22]$
and $\ell \ge |m|$.  We now introduce a coordinate defined by \be
dy^*=\frac{dy}{y(y+4\pi \tilde{T}_R)}, \ee which is integrated to be
\be \label{newyc}y^*=\frac{1}{4\pi
\tilde{T}_R}\ln\Big[\frac{y}{y+4\pi \tilde{T}_R}\Big].\ee This maps
$y\in [0,\infty)$ to $y^*\in(-\infty,0]$. Then, the equation
(\ref{rad1}) becomes the Schr\"{o}dinger-type equation
\begin{eqnarray} \label{ysol}
\frac{d^2Y}{dy^{*2}}+\Big(\tilde{\omega}^2-V_{\rm NEK}(y)\Big)Y=0.
\end{eqnarray}
The potential $V_{\rm NEK}(y)$ is given by
\begin{eqnarray}\label{nekv}
V_{\rm NEK}(y)=\tilde{\omega}^2+(\bar{K}_{\ell m}-m^2)y(y+4\pi
\tilde{T}_R)-\Big(\tilde{\omega}+m(y+2\pi\tilde{ T}_R)\Big)^2,
\end{eqnarray}
which implies that $\lim_{y\to \infty}V_{\rm NEK} \to \infty$ for
$\ell>|m|$. This reflects the nature of AdS$_2$ base in
(\ref{nhek}). In other words, the case of $\ell=m$ is not allowed
for our computation.

 On the other hand, in the near-horizon limit, it takes
the form \be V_{\rm
NEK+}(y)~\approx~\tilde{\omega}^2-(\tilde{\omega}+2\pi \tilde{T}_R
m)^2, \ee which leads to zero for $m=0$. In this case, a solution to
(\ref{ysol}) is given by \be
\label{newysol}Y_+(y^*)=Ae^{-i(\tilde{\omega}+2\pi \tilde{T}_R
m)y^*}+Be^{i(\tilde{\omega}+2\pi \tilde{T}_R m)y^*}, \ee where the
first term (the second one) correspond to the ingoing mode (outgoing
mode) when considering (\ref{newan}).
 We are now in a position to find QNFs of scalar field
around  the black hole  (\ref{nhek}). It turns out that the solution
to Eq.(\ref{rad1}) is given by the hypergeometric functions as
\begin{eqnarray}
Y(y)&=&c_1y^{-\frac{i}{2}\left(m+\frac{\tilde{\omega}}{2\pi
\tilde{T}_R}\right)}(y+4\pi
\tilde{T}_R)^{-\frac{i}{2}\left(m-\frac{\tilde{\omega}}{2\pi
\tilde{T}_R}\right)}\times\nn\\
&&{}_2F_1\Big[-im+\frac{1}{2}+\beta,~-im+\frac{1}{2}-\beta,
~1-i\left(m+\frac{\tilde{\omega}}{2\pi
\tilde{T}_R}\right),~-\frac{y}{4\pi \tilde{T}_R}\Big]\nn\\
&&\hspace*{-1em}+~c_2y^{\frac{i}{2}\left(m+\frac{\tilde{\omega}}{2\pi
\tilde{T}_R}\right)}(y+4\pi
\tilde{T}_R)^{-\frac{i}{2}\left(m-\frac{\tilde{\omega}}{2\pi
\tilde{T}_R}\right)}\times\nn\\
&&{}_2F_1\Big[\frac{i\tilde{\omega}}{2\pi
\tilde{T}_R}+\frac{1}{2}+\beta,~\frac{i\tilde{\omega}}{2\pi
\tilde{T}_R}+\frac{1}{2}-\beta,
~1+i\left(m+\frac{\tilde{\omega}}{2\pi\tilde{
T}_R}\right),~-\frac{y}{4\pi \tilde{T}_R}\Big]
\end{eqnarray}
where $c_{1,2}$ are arbitrary constants and $\beta$ is given by
\begin{eqnarray}
\beta=\sqrt{\frac{1}{4}+\bar{K}_{\ell m}-2m^2}.
\end{eqnarray}
We note that the two first terms ($c_1 y^\cdot,c_2y^\cdot$) can be
recovered from considering (\ref{newysol}) together with
(\ref{newyc}).

 In order to obtain
QNFs, we first require the ingoing mode at horizon and then,
Dirichlet boundary condition at infinity\footnote{The near-horizon
geometry of NEK black hole we consider in this section admits the
potential behavior as $V_{\rm NEK}\to0$ at horizon and $V_{\rm
NEK}\to\infty$ in the $y\to\infty$ limit as given in
Eq.(\ref{nekv}), which  suggests that its asymptote is changed from
a flat spacetime given by the Kerr black hole to an AdS spacetime.
This means that QNFs of scalar field propagating around the
near-horizon geometry of NEK black hole can be defined not by the
outgoing mode, but by $Y=0$ for the Dirichlet boundary condition at
asymptotically AdS infinity. We state  clearly that  these QNFs can
not be consistent with those obtained in a {\it whole} geometry of
NEK black hole \cite{Hod:2008zz}. Nevertheless, in appropriate
limits, two QNFs can be matched (see Table 1 in Sec.6 for detailed
limits), which is one of our main results in this paper.}. Choosing
the ingoing mode at horizon $(y=0)$ leads to $c_2=0$ and then, the
solution at infinity ($y \to \infty$) can be written as
\begin{eqnarray}
Y(y)=\Gamma_1y^{-\frac{1}{2}-\beta}+\Gamma_2y^{-\frac{1}{2}+\beta},
\end{eqnarray}
where $\Gamma_{1,2}$ are given by
\begin{eqnarray}
\Gamma_1&=&c_1\frac{\Gamma(-2\beta)\Gamma\Big[1-i(m+\frac{\tilde{\omega}}{2\pi
T_R})\Big]}
{\Gamma(-i m+\frac{1}{2}-\beta)\Gamma(-i\frac{\tilde{\omega}}{2\pi \tilde{T}_R}+\frac{1}{2}-\beta)},\\
\Gamma_2&=&c_1\frac{\Gamma(2\beta)\Gamma\Big[1-i(m+\frac{\tilde{\omega}}{2\pi
\tilde{T}_R})\Big]} {\Gamma(-i
m+\frac{1}{2}+\beta)\Gamma(-i\frac{\tilde{\omega}}{2\pi
T_R}+\frac{1}{2}+\beta)}.
\end{eqnarray}
Imposing the Dirichlet condition at infinity leads to $\Gamma_2=0$,
which provides two conditions
\begin{eqnarray}
\label{qnf1}&&-i m+\frac{1}{2}+\beta=-n,\\
\label{qnf2}&&-i\frac{\tilde{\omega}}{2\pi
\tilde{T}_R}+\frac{1}{2}+\beta=-n.
\end{eqnarray}
From (\ref{qnf2}), one finds the purely imaginary QNFs
\begin{eqnarray}
\tilde{\omega}=-i2\pi\tilde{ T}_R\Big(n+\frac{1}{2}+\beta\Big).
\end{eqnarray}
Further, considering the transformation (\ref{ctran}), we have the
relation of time and angular-part \be e^{-i\omega
t+im\phi}=e^{-i\tau[\frac{2M}{\lambda}\omega-\frac{m}{\lambda}]+im\varphi}=e^{-i\tau
\tilde{\omega}+im\varphi}, \ee which implies that QNFs are given by
\be \label{2dqnf}\omega=m\Omega -i 2\pi T_{\rm
NEK}\Big(n+\frac{1}{2}+\beta\Big). \ee Also, we have \be
\omega_R=\omega,~~T_R=2MT_{\rm NEK}. \ee From (\ref{qnf1}), one has
\begin{eqnarray}
m=-i2\pi T_L\Big(n+\frac{1}{2}+\beta\Big),
\end{eqnarray}
which implies that azimuthal number $m$ is purely imaginary for real
$\beta>0$.

Finally, we  wish to mention that for $\ell \gg m$, $\beta$ is
approximated to be \be \beta=\sqrt{\ell(\ell+1)+1/4-(2-c)m^2}\approx
\ell+\frac{1}{2},\ee which allows us  to rewrite $\omega$ as \be
\label{2dapp} \omega \approx m\Omega -i 2\pi T_{\rm
NEK}\Big(n+\ell+1\Big), \ee being consistent with the result
(\ref{impq}) for a nearly extremal case of replacing $T_H$ by
$T_{\rm NEK}$.

\newpage
\section{Summary and conclusion}

\begin{table*}[t]
\begin{center}
\begin{tabular}{|c|c|c|c|}
\hline \hline
 QNFs & Eq. (4.28) & Eq. (5.25) & Eq. (6.19) \\
\hline background&  & near-horizon geometry & \\
geometry  & Kerr & of NEK & NEK\\
      &  & (asymptotically AdS) & (asymptotically flat) \\
\hline how to  &  &
boundary conditions: & boundary conditions:\\
find & operator  & ingoing mode (near horizon)
& ingoing mode (near horizon)\\
 QNFs & method & ~~~~~~~~0 ~~~~~~(at infinity) & outgoing mode (at infinity)\\
\hline permitted & & & non-equatorial:\\
 QN modes & all & $\ell\gg m$ & $\ell\neq m \geq 0$\\
\hline taking   & near-horizon & & \\
the limit & near-region & no limit & near-region \\
\hline further &nearly&&\\
restriction & extremal&no further restriction&no further restriction\\
\hline final& \multicolumn{3}{c|}{}\\
QNFs &\multicolumn{3}{c|}
{$\omega \simeq m\Omega-i2\pi T_{\rm NEK}(n+\ell+1)$~~~~~~}\\
\hline
\end{tabular}
\end{center}
\caption[crit]{Comparison between the QNFs approaches}
\end{table*}

In this paper, we have obtained two analytic expressions for QNFs:\\
\hspace*{0.8em}(i) ~(\ref{impq}), found by using the hidden
conformal
symmetry developed in the near-horizon,\\
\hspace*{0.8em}(ii) (\ref{2dqnf}), obtained considering the
near-horizon geometry of the NEK black hole. \\
The reference QNFs (\ref{aqnfse}) was shown in Appendix B, which are
those of non-equatorial ($\ell\not=m$) scalar modes around the NEK
black hole. They are different apparently, but it was shown that
QNFs (i) correspond approximately to not only those (\ref{2dapp}) of
scalar perturbation around the near-horizon region of NEK black
hole, but also those (\ref{aqnfse}) of non-equatorial ($\ell\not=
m$) scalar modes around the NEK black hole.

On the other hand, the hidden conformal symmetry approach based on
(\ref{oh1})-(\ref{obhm}) with the potential (\ref{potss}) in the
{\it near-region} could not determine QNFs of scalar around the NEK
black hole, while QNFs (i) indicates that the hidden conformal
symmetry by using (\ref{rrhoeq}) given in the {\it near-horizon
region} as well as {\it near-region} provides the same as QNFs
(\ref{2dapp}) for the nearly extremal case.

We mention that the approaches (i) and (ii) do not yield QNFs
(\ref{eqqnf}) of equatorial $(\ell=m$) modes around the NEK black
hole. This is because the hidden conformal symmetry  and
near-horizon geometry approaches  keep the near-horizon feature of
the scalar potential around the NEK black hole only (see Table 1).
It should be pointed out that $\delta^2$ (\ref{delt}) always take a
negative value in the near-region and low-frequency limits, which
implies that the equatorial mode ($\delta^2>0$) and near-region
limit are incompatible to each other. That is why
 the approaches (i) and (ii), obtained from the near-region and near
 horizon,
do not admit the long-lived equatorial mode.

Finally, we conclude that the hidden Kerr/CFT proposal could
determine quasinormal modes and frequencies of the massless scalar
wave propagating around the NEK (rapidly rotating) black hole
partly.

\vspace{1cm}
\section*{Acknowledgement}
  This  was  supported
by the National Research Foundation of Korea (NRF) grant funded by
the Korea government (MEST) (No.2012-R1A1A2A10040499). Y.M. was
supported partly by the National Research Foundation of Korea (NRF)
grant funded by the Korea government (MEST) through the Center for
Quantum Spacetime (CQUeST) of Sogang University with grant number
2005-0049409.

\newpage
\section*{Appendix A: Explicit derivation of (\ref{vkom})}
We first arrange the potential (\ref{vrho}) as
\begin{eqnarray}
V_{\rm K}(\rho)&=&\omega^2-\frac{\kappa^2}{\sinh^2[\kappa\rho]}
\tilde{K}_{\ell}(\rho)\nonumber\\
&=&\omega^2-\Bigg[\frac{\kappa}{r_0}\Big((M+r_0\coth[\kappa\rho])^2+a^2\Big)\omega
-\frac{\kappa}{r_0}am\Bigg]^2\nonumber\\
&&\hspace*{11em}-~\frac{2\kappa^2ma\omega}{\sinh^2[\kappa\rho]}
+\frac{\kappa^2a^2\omega^2}{\sinh^2[\kappa\rho]}+\frac{\kappa^2K_{\ell}}{\sinh^2[\kappa\rho]}.
\label{apa1}
\end{eqnarray}
In the low frequency ($\omega a,~\omega M\ll1$) and near-horizon
limits,  the second line of Eq. (\ref{apa1}) becomes
\begin{eqnarray}
&&\omega^2-\Bigg[\frac{\kappa}{r_0}\Big((M+r_0\coth[\kappa\rho])^2+a^2\Big)\omega
-\frac{\kappa}{r_0}am\Bigg]^2\nonumber\\
&\Rightarrow&\omega^2-\Bigg[\frac{\kappa}{r_0}\Big((M+r_0)^2+a^2\Big)\omega
-\frac{\kappa}{r_0}am\Bigg]^2\nonumber\\
&=&\omega^2-\Bigg[\frac{\kappa}{r_0}\Big(r_+^2+a^2\Big)\omega
-\frac{\kappa}{r_0}am\Bigg]^2\label{se1}\\
&=&\omega^2-(\omega-m\Omega)^2\label{se2},
\end{eqnarray}
where we used $r_+=M+r_0$ in (\ref{se1}) and
$\kappa=r_0/(r_+^2+a^2)=r_0\Omega/a$ in (\ref{se2}). A dominant term
in the third line of Eq. (\ref{apa1}) is given by
\begin{eqnarray}
\frac{\kappa^2K_{\ell}}{\sinh^2[\kappa\rho]}
=\frac{\ell(\ell+1)\kappa^2}{\sinh^2[\kappa\rho]}.\label{se4}
\end{eqnarray}
From (\ref{se2}) and (\ref{se4}), we finally obtain (\ref{vkom}).

\newpage
\section*{Appendix B: QNFs of NEK black holes}

In this Appendix we briefly show  an analytic form of QNFs of scalar
around NEK black holes in asymptotically flat
spacetime~\cite{Cardoso:2004hh,Hod:2008zz}. To this end, we first
recall the radial equation (\ref{oradial}) which is reexpressed  in
terms of $k=r^2\omega+a^2\omega-am$ as
\begin{eqnarray}\label{ap1}
\triangle\frac{\partial^2 R}{\partial r^2}+2(r-M)\frac{\partial
R}{\partial
r}+\left(\frac{k^2}{\triangle}+2ma\omega-a^2\omega^2-K_{\ell}\right)R=0.
\end{eqnarray}
 Introducing new dimensionless variables
\begin{eqnarray}
x=\frac{r-r_+}{r_+},~~~\bar{\tau}=\frac{2r_0}{r_+},
~~~\bar{\omega}=\frac{\omega-m\Omega}{2\pi
T_{H}},~~~\hat{\omega}=\omega r_+,
\end{eqnarray}
(\ref{ap1}) can be rewritten as
\begin{eqnarray}\label{main}
x(x+\bar{\tau})R^{\prime\prime}+(2x+\bar{\tau})R^{\prime}+WR=0,
\end{eqnarray}
where the prime (${}^{\prime})$ denotes differentiation with respect
to $x$ and  $W$ is
\begin{eqnarray}
W=\frac{k^2}{x(x+\bar{\tau})r_+^2}+2ma\omega-a^2\omega^2-K_{\ell}.
\end{eqnarray}
We mention that the double limit of  $a\to M$ and $\omega\to
m\Omega$ correspond to $\bar{\tau}\to0$ and $\bar\omega\to 0$. The
former limit  indicates the NEK black hole, while the latter shows
that $\omega_R$ is almost given by $m\Omega$ in the NEK black hole.
Note also that for QNMs, two boundary conditions near the horizon
and at infinity are given by
\begin{eqnarray}\label{bc}
R\sim\left\{\begin{array}{ll} (i)~~{\rm~ingoing~
waves~}\hspace*{2.55em} {\rm as}~~x\to0~~ (r\to r_+),
\label{r1}\\
(ii)~{\rm~outgoing~ waves~}\hspace*{2em} {\rm as}~~x\to \infty~~
(r\to \infty).\label{r2}\end{array}\right.
\end{eqnarray}
In the far region ($x\gg {\rm max}\{\bar\tau,\bar\omega\}$),
(\ref{main}) becomes
\begin{eqnarray}\label{far}
x^2R^{\prime\prime}+2xR^{\prime}+W_{\rm far}R=0,
\end{eqnarray}
where
\begin{eqnarray}
W_{\rm far}=\omega^2r_+^2x^2+4\omega\hat{\omega}r_+x+
4\hat{\omega}^2+2ma\omega-a^2\omega^2-K_{\ell}.
\end{eqnarray}
A solution to (\ref{far}) is expressed  in terms of the confluent
hypergeometric function,
\begin{eqnarray}
\label{as1} R_{\rm far}&=&A_1 (2i\omega
r_+)^{\frac{1}{2}+i\delta}x^{-\frac{1}{2}+i\delta}e^{-i\omega
r_+x}~U(i\delta+2i\hat{\omega}+1/2,~1+2i\delta,~2i\omega
r_+x)\nn\\
&&\hspace*{-2em}+A_2 (2i\omega
r_+)^{\frac{1}{2}-i\delta}x^{-\frac{1}{2}-i\delta}e^{-i\omega
r_+x}~U(-i\delta+2i\hat{\omega}+1/2,~1-2i\delta,~2i\omega r_+x),
\end{eqnarray}
where an important quantity $\delta$ is given by
\begin{eqnarray}\label{delt}
\delta^2=4\hat\omega^2+2ma\omega-a^2\omega^2-K_{\ell}-\frac{1}{4},
\end{eqnarray}
which is either positive  or negative. Here $K_\ell$ is given by \be
K_\ell =\ell(\ell+1)+{\cal O}(a\omega). \ee

On the other hand,  in near-horizon region ($x\ll1$), (\ref{main})
reduces to
\begin{eqnarray}\label{near}
x(x+\bar{\tau})R^{\prime\prime}+(2x+\bar{\tau})R^{\prime}+W_{\rm
near}R=0,
\end{eqnarray}
where $W_{\rm near}$ is
\begin{eqnarray}
W_{\rm near}=\frac{(2\hat\omega
x+\bar\omega\bar\tau/2)^2}{x(x+\bar{\tau})}+2ma\omega-a^2\omega^2-K_{\ell}.
\end{eqnarray}
Imposing the ingoing boundary condition  $(i)$ in (\ref{bc}), one
finds a solution to (\ref{near}) as
\begin{eqnarray}\label{as2}
\hspace*{-1em}R_{\rm near}=x^{-\frac{i}{2}\bar\omega}
\left(\frac{x}{\bar\tau}+1\right)^{i(\bar\omega/2-2\hat\omega)}
{}_2F_1(i\delta-2i\hat{\omega}+1/2,~-i\delta-2i\hat{\omega}+1/2
~1-i\bar\omega,~-x/\bar\tau),
\end{eqnarray}
where ${}_2F_1$ is the hypergeometric function. A remaining  task is
to match  solution  (\ref{as1}) with (\ref{as2}) in the overlapping
region ${\rm max}\{\bar\tau,\bar\omega\}\ll x\ll 1$.  It turns out
that this matching condition determines  $A_1$ and $A_2$ to be
\begin{eqnarray}\label{a1}
A_1&=&\bar\tau^{-\frac{i}{2}\bar\omega+\frac{1}{2}-i\delta}(2i\omega
r_+)^{-\frac{1}{2}-i\delta}
\frac{\Gamma(2i\delta)\Gamma(1-i\bar{\omega})}{\Gamma(i\delta+\frac{1}{2}-2i\hat\omega)
\Gamma(i\delta+\frac{1}{2}+2i\hat\omega-i\bar{\omega})},\\
A_2&=&\bar\tau^{-\frac{i}{2}\bar\omega+\frac{1}{2}+i\delta}(2i\omega
r_+)^{-\frac{1}{2}+i\delta}
\frac{\Gamma(-2i\delta)\Gamma(1-i\bar{\omega})}{\Gamma(-i\delta+\frac{1}{2}-2i\hat\omega)
\Gamma(-i\delta+\frac{1}{2}+2i\hat\omega-i\bar{\omega})}.\label{a2}
\end{eqnarray}
Substituting (\ref{a1}) and (\ref{a2}) into (\ref{as1}), and
imposing the outgoing boundary condition $(ii)$ in (\ref{bc}), we
get the quasinormal condition
\begin{eqnarray}
&&\hspace*{-2em}\frac{\Gamma(2i\delta)\Gamma(1+2i\delta)(-2i\bar\tau\hat\omega)^{-i\delta}}
{[\Gamma(i\delta+\frac{1}{2}-2i\hat\omega)]^2
\Gamma(i\delta+\frac{1}{2}-i\bar\omega+2i\hat\omega)}=-
\frac{\Gamma(-2i\delta)\Gamma(1-2i\delta)(-2i\bar\tau\hat\omega)^{i\delta}}
{[\Gamma(-i\delta+\frac{1}{2}-2i\hat\omega)]^2
\Gamma(-i\delta+\frac{1}{2}-i\bar\omega+2i\hat\omega)}.\nn
\end{eqnarray}
After manipulations with taking the double limits of $\bar\tau,
\bar\omega\ll1$, we find the QNFs of scalar around NEK black hole
for non-equatorial ( $\ell\neq m\geq0$) modes with
$\delta^2<0$~\cite{Hod:2008zz}
\begin{eqnarray} \label{aqnfs}
\omega\simeq m\Omega-i2\pi T_{\rm
NEK}\Big[n+\frac{1}{2}-i\delta\Big],~~\Omega=\frac{1}{2M},~~T_{\rm
NEK}=\frac{r_0}{4\pi M^2} \ll 1.
\end{eqnarray}
 In the low-frequency limit of $\omega a\le\omega M\ll 1$ and $\hat{\omega}= \omega r_+ \ll 1$ for the NEK black hole,
$\delta^2$ is negative as \be \delta^2 \simeq
-\frac{1}{4}-\ell(\ell+1) \ee
 and thus, $\delta$ is given by  \be
\delta=i\Big(\ell+\frac{1}{2}\Big).\ee
 Plugging $\delta$ into
(\ref{aqnfs}) leads to \be
 \label{aqnfse}
\omega \simeq m\Omega-i2\pi T_{\rm NEK}\Big(n+\ell+1\Big), \ee which
is the same form as in (\ref{impq}) for a nearly extremal case and
as in (\ref{2dapp}).

On the other hand, for equatorial $\ell=m\geq0$ modes with
$\delta^2>0$, their  QNFs take an approximate form
\begin{eqnarray} \label{eqqnf}
\omega \simeq m\Omega-i2\pi T_{\rm NEK}\Big(n+\frac{1}{2}\Big),
\end{eqnarray}
which implies that non-equatorial modes decay faster than equatorial
modes.

\end{document}